# Coloration de nombre de Grundy pour les graphes triangulés

Ali Mansouri[*], Mohamed Salim bouhlel[**]

mehermansouri@yahoo.fr

medsalim.bouhlel@enis.rnu.tn

**Abstract:** Notre travail s'intègre dans la problématique générale de la stabilité du réseau ad hoc. Plusieurs, travaux ont attaqué ce problème. Parmi ces travaux, on trouve la modélisation du réseau ad hoc sous forme d'un graphe (les machines correspondent aux nœuds, les arrêtes correspondent aux liens entre les machines).

Donc le problème de stabilité du réseau ad hoc qui correspond à un problème d'allocation de fréquence se résume à un problème d'allocation des couleurs aux nœuds de graphe. Souvent les transmetteurs sont disposés comme les sommets d'un réseau triangulaire dans le plan. Ce modèle est souvent utilisé car il offre une bonne couverture pour le réseau.

Dans cet article, nous présentons un algorithme de coloration maximale des graphes triangulés en utilisant un paramètre de coloration « le nombre de Grundy ».

**Mots clés:** réseau ad hoc, modélisation du réseau ad hoc, graphe, stabilité du réseau, nombre de Grundy.

## INTRODUCTION

Les réseaux ad hoc sont parfois définis comme des réseaux spontanés sans fil [ABL94]. Ils réunissent un grand nombre d'objets communicants sans fil, sans infrastructure et tous ces objets peuvent se déplacer. De tels réseaux sont donc intrinsèquement différents des réseaux classiques qui utilisent une dorsale filaire et des collecteurs de trafic pour connecter plusieurs réseaux locaux filaires ou sans fil. Les réseaux ad hoc doivent s'auto-organiser pour acheminer le trafic d'un point à l'autre du réseau ad hoc. L'auto organisation passe d'abord par une solution d'acheminement du trafic, puisque la source et la destination peuvent ne pas être à portée radio. Le réseau doit donc collaborer avec de potentiels nœuds intermédiaires, s'auto attribuer des adresses... Toutes les fonctionnalités doivent à terme se déployer automatiquement sans paramétrage éventuel de l'utilisateur.

Il s'avère donc, que le problème étant d'en allouer le nombre minimum, la valeur elle-même des fréquences est sans importance. Ce que les théoriciens des graphes ont pris l'habitude de faire est tout simplement de remplacer la valeur numérique d'une fréquence par une couleur. Il suffit alors d'affecter à chaque sommet du graphe une couleur en faisant bien attention à ce que deux sommets adjacents, c'est-à-dire joints par un arc, ne possèdent pas la même couleur. On appelle ça une coloration du graphe. [AKR00] Le problème d'allocation de fréquences se résume donc à un problème de coloration de graphe utilisant le nombre minimum de couleurs possibles. [AMC94].

Plusieurs travaux ont attaqués la coloration de graphe. B. Levêque et F. Maffray [BLM04] ont proposé un algorithme pour la coloration des graphes en un temps linéaire.

Aussi, F. Gavril, [FGA72] a proposé un algorithme pour une coloration minimum de graphes. R. W. Irving et D. F. Manlove [RWM99] ont proposé un algorithme de coloration des graphes avec le nombre chromatique.

Un autre travail de coloration d'arbres est proposé par S. Hedetniemi, and T. Beyer. [SHB82] qui utilise un autre paramètre de coloration « le nombre de Grundy »

Dans ce contexte, nous proposons un algorithme de coloration de graphes triangulés puisque les transmetteurs sont souvent disposés comme les sommets d'un réseau triangulaire dans le plan. Ce modèle est souvent utilisé car il offre une bonne couverture pour le réseau.

Notre algorithme de coloration se base sur les propriétés des graphes triangulés et utilise le nombre de Grundy comme paramètre de coloration.

Nous présentons cet article de la manière suivante:

Dans la section 2, nous présentons quelques définitions de base, ensuite dans la section 3 nous proposons notre algorithme de coloration, dans la section 4 nous présentons l'utilité de notre algorithme et nous finirons par une conclusion et quelques perspectives.

# 1. Preliminaries

Les réseaux ad hoc sont modélisés par des graphes. Un graphe est défini par la liste de ses sommets (les points) et de ses arêtes (les lignes existant entre certains couples de points). Un graphe G est défini aussi par un ensemble de sommets V (G) et un ensemble d'arêtes E (G).

Deux sommets sont adjacents, s'ils sont reliées par une arrête.

## 1.1. Definition d'un graphe triangulé

Un graphe est dit triangulé s'il ne contient aucun cycle induit de longueur supérieure ou égale à quatre (les graphes triangulés apparaissent sous le nom de chordal graphe dans la littérature anglophone).

Un graphe G est dit faiblement triangulé si ni G, ni G', ne contiennent de cycles induits de longueur au moins 5. M. Burlet [MBU81] montre que ces graphes sont parfaits et donne un algorithme polynomial pour les reconnaître. Les problèmes d'optimisation dans cette classe sont résolus par F. Maire [FMA93] ainsi que par M. Burlet and J.-P. Uhry [MBU84].

Les graphes triangulés correspondent exactement aux graphes d'intersection des sous arbres dans un arbre, ce qui leur confère certaines applications dans le domaine de la classification. Ils sont reconnaissables en temps et espace linéaire et les problèmes de la clique maximum, de la coloration minimum, du stable maximum et de la partition minimum en cliques peuvent être résolus en temps et espace linéaire lorsque l'on se restreint à ceux-ci.

Une sous-classe intéressante des graphes triangulés est la classe des graphes scindés.

Un graphe scindé est un graphe dont les sommets admettent une partition en deux sous-ensembles S et C où S un est stable et C une clique ; par analogie aux graphes bipartis, nous noterons G = (S, C, E) le graphe en question. Clairement, le complément d'un graphe scindé est aussi un graphe scindé ; en fait, il s'avère qu'un graphe G est scindé si et seulement si G et G' sont triangulés

## 1.2. Propriétés des graphes triangulés

Ce type de graphe a plusieurs propriétés dont on peut citer :

### 1.2.1. *La notion de sommet complété (simplicial)*
**Définition:**

Un sommet d'un graphe est dit complété (en anglais « simplicial ») si son voisinage N(x) est une clique.

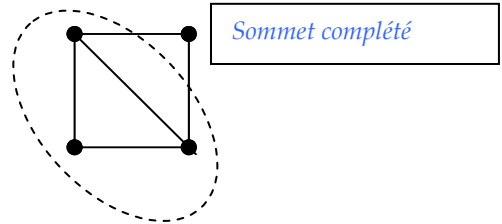

*Figure 1: sommet completé*

**Théorème :** [ENG56]

Tout graphe triangulé admet un sommet complété. De plus, si le graphe n'est pas complet, il admet deux sommets complétés non voisins. (Contient au moins deux sommets complétés non adjacents).

### 1.2.2. *La notion de schéma d'élimination parfait*
**Définition** :

Etant donné un graphe G d'ordre n. {Vi ….vn} est un ordre total sur l'ensemble des sommets V (G), si i Є [1 ….    N], vi est un sommet complété du graphe induit par   {v1, v2, v3,…..…vn}

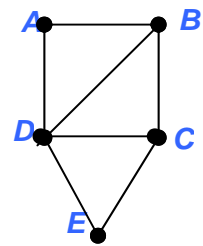

*Figure 2: schema d'elimination parfait*

On donne le schéma d'élimination parfait pour l'exemple suivant :

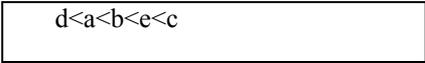

d<a<b<e<c

**Théorème :** [ENG56]

Un graphe est triangulé si et seulement s'il possède un schéma d'élimination parfait.

Par la suite, on donne la définition d'une coloration et le nombre de Grundy

On appelle une coloration du graphe, l'affectation à chaque sommet du graphe une couleur en faisant bien attention à ce que deux sommets adjacents, c'est-à-dire joints par un arc, ne possèdent pas la même couleur.

En effet, il ya plusieurs travaux de coloration de graphes, par exemple on trouve la coloration des sommets, la coloration des arrêtes [BEH03]. [FGA72]

## 1.3. Definition d'une coloration

La coloration d'un graphe nécessite l'utilisation de plusieurs paramètres de coloration. Dans la littérature, on trouve plusieurs paramètres fondamentaux comme le nombre chromatique, le Grundy partiel, le Grundy total.

## 1.4. definition du nombre chromatique

Le nombre chromatique est un paramètre de coloration de graphe, qui utilise le minimum de couleurs. La détermination du nombre chromatique d'un graphe est dans le cas général un problème NP-difficiles. De nombreux travaux ont donc été menés pour définir des bornes pour ce paramètre en fonction d'autres paramètres de graphe. Les auteurs montrent dans [FHW96] plusieurs relations entre le nombre chromatique d'un graphe et des paramètres de domination de ce graphe.

Beaucoup d'autres types de graphes ont été étudiés pour ce paramètre. Nous pouvons citer par exemple les graphes circulants G (n, S), qui sont des graphes d'ordre n où S est l'ensemble des distances dans le graphe. *Hedetniemi,* [PHG03], en 2003, fournit entre autre la valeur du nombre chromatique pour des graphes circulants.

### 1.4.1. *Limites du nombre chromatique :*

Le paramètre utilisé « le nombre chromatique » pose des problèmes. En effet, ce paramètre ne permet pas l'ordonnancement des événements de même durés qui peuvent se dérouler en même temps. Donc plusieurs événements de mêmes couleurs peuvent se dérouler en même temps, d'où le problème d'allocation de fréquences qui se résume à un problème de coloration de graphe, utilisant le nombre minimum de couleurs possibles, n'est pas encore résolu.

Dans notre étude nous s'intéressons à l'étude de nombre de Grundy qui est utilisé dans notre algorithme de coloration.

## 1.5. Le nombre de Grundy

Dans cette partie, nous allons étudier un second paramètre maximisant le nombre de couleurs nécessaires à la coloration d'un graphe.

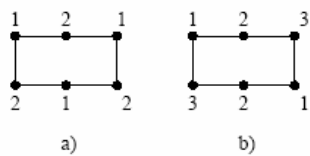

*Figure 3 : coloration de Grundy*

Nous pouvons noter qu'une coloration de Grundy d'un graphe peut être obtenue par un algorithme "glouton" de recherche du nombre chromatique du graphe. En effet, l'algorithme va chercher toutes les colorations possibles des sommets du graphe. Le nombre chromatique sera alors donné par la coloration qui minimise le nombre de couleurs utilisées et le nombre de Grundy sera donc donné par celle qui maximise le nombre de couleurs.

Ce paramètre a été introduit par C. Germain and H. Kheddouci [CGH02]. Ils ont prouvé que la détermination du nombre de Grundy pour un graphe quelconque était un problème

On peut voir que le nombre entier minimum pour lequel un graphe G a une k-coloration de Grundy est toujours égal au nombre chromatique $\gamma$ (G). ($\gamma$ (G) $\leq$ $\Gamma$(G)).

### 1.5.1. *Définition*

Une k-coloration de Grundy est une k-coloration propre vérifiant la propriété suivante : chaque sommet v, coloré par, une couleur i (avec $1 \leq i \leq k$), doit être adjacent à au moins $i-1$ sommets colorés par chacune des couleurs j telles que $1 \leq j \leq i-1$. Le nombre de Grundy $\Gamma$ (G) d'un graphe G est le nombre maximum de couleurs nécessaires pour avoir une coloration de Grundy de G.

### 1.5.2. *Propriétés de nombre de Grundy*

De la définition d'une coloration de Grundy, nous pouvons déduire la propriété suivante :

**Soit G un graphe, alors $\Gamma$ (G) $\leq \Delta + 1$.**

**Preuve.** D'après la construction d'une coloration de Grundy, chaque sommet $x \in G$ de couleur i doit être adjacent aux couleurs 1, 2, ..., i-1. Ainsi nous avons :

$$c(x) \leq d(x) + 1 \leq \Delta(G) + 1.$$

Nous en déduisons donc :

$$\Gamma(G) \leq \Delta(G) + 1$$

Les auteurs ont donné un algorithme linéaire pour déterminer le nombre de Grundy d'un arbre et ont montré une relation d'ordre entre le nombre de Grundy, le nombre chromatique et le nombre achromatique pour tout graphe G : [CHS79]

$$\chi(G) \leq \Gamma(G) \leq \psi(G).$$

En 1997, les auteurs ont donné un algorithme pour le nombre de Grundy pour les k-arbres partiels [CHS79]

**Théorème 5 :** Soit G un k-arbre partiel d'ordre n, son nombre de Grundy peut être calculé en O (n3K).

**Théorème 6 :** Le nombre de Grundy d'un k-arbre partiel G d'ordre n, $n \geq k \geq 1$, est au plus $1 + k \log_2 n$.

Germain et Kheddouci [CGH02] ont étudié le nombre de Grundy de graphes puissances. Ils ont donné des bornes de ce paramètre pour les graphes puissances d'une chaîne, d'un cycle, d'une chenille et d'un arbre binaire complet.

Ils ont montré que le nombre de Grundy est croissant pour l'ordre des sous graphes induits :

**Théorème 7** : Soit G = (V, E) un graphe et V' ⊆ V. s'il existe une k-coloration de Grundy pour le sous graphe induit donné par V', alors :

$$\Gamma(G) \geq K.$$

Dans le théorème suivant ils ont travaillé sur les graphes ayant un ensemble stable de sommets.

**Théorème 8** : [FRI99] Soit G un graphe connecté d'ordre n et de nombre de stabilité α. Alors :

$$\Gamma(G) \leq n + 1 - \alpha.$$

En 1956, les auteurs [16] ont établi plusieurs inégalités pour les nombres chromatiques χ et χ̄ d'un graphe G = (V, E) et de son complément Ḡ, dont

$$2\sqrt{n} \leq \chi + \bar{\chi} \leq n + 1 \text{ où } n = |V|.$$

La borne supérieure de cette relation a été étudiée pour de nombreux paramètres : le nombre chromatique **[FMA94]**, des paramètres de domination et d'autres paramètres. Pour notre part, nous montrons cette borne pour le nombre de Grundy.

**Théorème 9 :** Soit G un graphe d'ordre n et Ḡ son complément. Soient Γ et Γ̄ les nombres de Grundy respectivement de G et de Ḡ. Alors

$$\Gamma + \bar{\Gamma} \leq n + 1$$

Par la suite nous présentons notre algorithme de coloration des graphes triangulés en utilisant le nombre de Grundy

## 2. Proposition d'un algorithme de coloration des graphes triangulés

### 2.1. Principe

Notre algorithme se compose de deux grandes étapes :

**1)** La détermination de schéma d'élimination parfait

**2)** La coloration du graphe en fonction de schéma d'élimination parfait

Colorer le premier sommet du schéma d'élimination parfait par le minimum de couleur

Pour le deuxième sommet, donner une couleur minimale qui n'est pas encore utilisée par ses voisins.

**3)** Tant qu'il y a des sommets a colorés, répéter l'étape 2.

### 2.2. Exemple :

*$1^{ère}$ étape : La détermination de schéma d'élimination parfait :*

Le schéma d'élimination parfait est obtenu en calculant tous les sommets complétés du graphe (voir exemple).

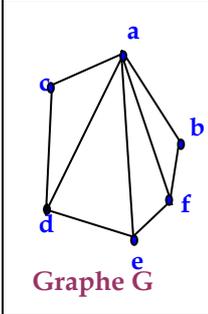

| 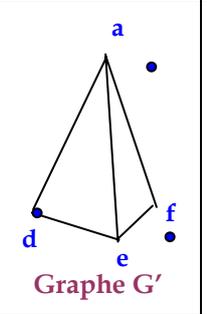 Graphe G | 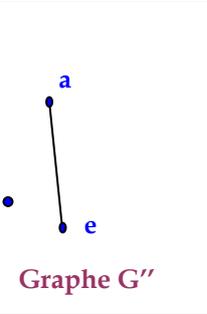 Graphe G' | Graphe G'' |
|---|---|---|
| Sommets complétés : b, c | Sommets complétés : d, f | Sommet complété : a, e |
| 1ère Partie Schéma d'élimination parfait : {b, c} | 2ème Partie Schéma d'élimination parfait : {b, c} < {d, f} | Schéma d'élimination parfait complet : {b, c} < {d, f} < {a, e} |

*Tableau 1 : composition du schéma d'élimination parfait d'un graphe*

*$2^{ème}$ étape : La coloration du graphe à partir du schéma d'élimination parfait :*

Nous allons maintenant construire une coloration de Grundy du graphe faiblement triangulé à partir de schéma d'élimination parfait.

Selon le schéma d'élimination parfait suivant {b, c} < {d, f} < {a, e}, nous allons construire une coloration du graphe G (a chaque étape de coloration du chaque sommet on vérifier la coloration de ses voisins).

**1)** Colorer le premier sommet du schéma d'élimination parfait par le minimum de la couleur (on affecte la couleur 1 au sommet b).

**2)** Pour le deuxième sommet donner une couleur minimale qui n'est pas encore utilisée par ses voisins.

**3)** tant qu'il y a des sommets a colorés, répéter l'étape 2.

Pour conclure on obtient enfin un nombre de Grundy égale à trois pour ce graphe.

### 2.3. Etude de la complexité de l'algorithme

La complexité de notre algorithme est la somme de complexité de deux algorithmes : algorithme de schéma d'élimination parfait et algorithme de coloration du graphe (puisque l'algorithme de coloration est basé sur l'algorithme de schéma d'élimination parfait).

La complexité de l'algorithme de schéma d'élimination est **n. $\Delta^2$**. En effet, pour un sommet A d'un graphe (G), on cherche son voisinage ($\Delta$) et le voisinage de son voisinage ($\Delta$). Donc, on répète le test pour les n sommet du graphe. Donc, la complexité est **n. $\Delta^2$**

Pour l'algorithme de coloration simple la complexité est **n. $\Delta$.** En effet, la coloration consiste à attribuer une couleur minimum au sommet spécifié, après la vérification de tous les sommets de son voisinage ($\Delta$). A chaque fois on vérifie le voisinage de sommet à colorer (**n**)

Donc, la complexité totale de l'algorithme est de **n.$\Delta^2$.**

De ce qui précède, on peut tirer le théorème suivant :

**Théorème 10:**

$$C(i) = \text{Min } \{q / q \geq 1 \text{ et } q \in Nc(i)\}$$

## 3. Utilité de l'algorithme proposé

On peut poser le problème d'assignation de fréquences dans un réseau radio ou de téléphones mobiles de la façon suivante :

Comment attribuer une fréquence à chaque émetteur du réseau, de telle manière que deux émetteurs qui peuvent interférer aient des fréquences ``suffisamment'' éloignées l'une de l'autre?

La contrainte d'optimisation est la suivante : on désire utiliser dans ce réseau une bande de fréquences la plus étroite possible (la bande passante étant une ressource rare). On modélise le réseau par un graphe et le problème est alors un problème de coloration d'un graphe.

Les sommets du graphe représentent les transmetteurs et deux sommets sont reliées par une arrête. Les fréquences affectées aux deux sommets adjacents doivent être différente.

Les sous graphes du réseau triangulaire est un schéma très utilisé car il offre la meilleure couverture avec des transmetteurs identiques. Notre algorithme permet de coloration des graphes triangulés, qui utilise un paramètre à savoir le « Grundy », permet d'assurer la stabilité du réseau a n'importe quel changement du réseau (ajout de lien, suppression de nœud...). Ceci nous permet de gérer à la fois la dynamicité du réseau et les défaillances transitoires est celle de l'auto-stabilisation, qui assure que le système converge vers une allocation de fréquence valide après une défaillance transitoire ou un changement de topologie. La base de notre algorithme consiste en une technique de regroupement rapide probabiliste, qui pourrait être exploitée afin de résoudre d'autres problèmes dans les réseaux de capteurs.

## 4. Conclusion

Les problèmes classiques d'optimisation (c.-a-d. ceux concernant les nombres chromatiques, de densité, de stabilisé et de couverture par des cliques) sont intéressants à plusieurs titres. Tout d'abord parce qu'ils apparaissent naturellement dans de nombreux problèmes pratiques et surtout lors de la stabilité de réseaux ad hoc.

Nous avons étudié, un paramètre de coloration pour lesquels nous cherchons à maximiser le nombre de couleurs nécessaires à la coloration d'un graphe.

Le nombre de Grundy permet, comme nous l'avons vu, de mettre en évidence la dominance de certains sommets d'un graphe.

Nous avons proposé un algorithme de coloration qui se repose sur des propriétés de graphes triangulés. Cet algorithme utilise la propriété de schéma d'élimination parfait

On a développé des algorithmes de coloration utilisant le maximum de couleur pour réduire le maximum les interférences dans le réseau et s'adaptent automatiquement, s'il y a des changements dans la topologie, pour maintenir une stabilité dans le réseau

Dans nos futurs travaux, nous cherchons à borner notre algorithme et l'appliquer sur plusieurs graphes et nous tentons aussi à développer des protocoles de communication qui seront également, implémentés et simuler par des scénarios bien élaborées et validées. Des comparaisons seront faites avec des protocoles existants.

**ACKNOWLEDGMENT**

Special think for Mr. Brice Effantin.

## 5. References

[Abl94] A. Blum. New approximation algorithms for graph coloring. *J ACM, 41 (3): 470-516, (1994)*.

[AKR00] A. Kapoor and R. Rizzi. Edge-coloring bipartite graphs. *Journal of Algorithms,34,390-396.2000*

[AMC94] A. McRae. NP-completness proof for Grundy coloring, 1994. *Unpublished manuscript,*

*personal communication*

[BEH03] B. Effantin and H. Kheddouci. The b-chromatic number of some power graphs. *Discrete Mathematics and Theoretical Computer Science, 6:45–54, 2003.*

[BLM04] B. Levêque et F. Maffray (2004). Coloring Meyniel graphs in linear time. *Cahier n° 105. Laboratoire Leibniz (IMAG), Grenoble, France.*

[CHS79] C. Christen and S. Selkow. Some perfect coloring properties of graphs. *Journal of Combinatorial Theory, B27:49–59, 1979*

[CGH02] C. Germain and H. Kheddouci. Grundy coloring for power graphs. *Electronic Notes in Discrete Mathematics Volume xx (Instructional Workshop and Symposium on Discrete Mathematics and Applications); submitted to Discrete Mathematics, 2002.*

[ENJ56] E. Nordhaus and J. Gaddum. On complementary graphs. *American Mathematical Monthly, 63:175–177, 1956.*

[FGA72] F. Gavril. Algorithms for minimum coloring, maximum clique, minimum covering by cliques and maximum independent set of a chordal graph. *SIAM J. Comput. , 1 (2): 180-187, (1972).*

[FHW96] F. Harary and T. W. Haynes. Nordhaus-Gaddum inequalities for domination in graphs. *Discrete Mathematics, 155:99–105, 1996.*

[FRI99] F. Roussel et I. Rusu (1999). Holes and dominoes in Meyniel graphs. *International Journal of Foundations of Computer Science 10, pp. 127 -146.*

[FMA93] F. Maire. Des polyominos aux graphes parfaits; contribution à l'étude des propriétés combinatoires de ces structures. PhD thesis, Université Paris 6, 1993

[FMA94] F. Maire. Slightly triangulated graphs are perfect. *Graphs and Combin., 10:263-268, 1994.*

[MBU81] M. Burlet. Étude algorithmique de certaines classes de graphes parfaits. PhD thesis, Université de Grenoble, 1981

[MBU84] M. Burlet and J.-P. Uhry. Parity graphs. In C. Berge and V. Chvátal, editors, *Topics on Perfect Graphs, Math. Stud. 88, pages 253-277, 1984.*

[PHG03] P. Erdos, S. T. Hedetniemi, R. C. Laskar and G. C. E. Prins. On the equality of the partial Grundy and upper ochromatic numbers of graphs. *Discrete Mathematics 272 (2003) 53-64.*

[RWM99] R. W. Irving and D. F. Manlove. The b-chromatic number of a graph. *Discrete appl. Math. , 91:127-141, (1999)*

[SHB82] S. Hedetniemi, S. Hedetniemi, and T. Beyer. A linear algorithm for the Grundy (coloring) number of a tree. Congressus Numerantium, 36:351–363, 1982.